\documentclass{article}
\usepackage{hyperref}
\usepackage{graphicx}
\usepackage[utf8]{inputenc}
\usepackage{color}

\usepackage{hyperref}

\makeatletter
\let\@fnsymbol\@arabic
\makeatother

\newcommand{\footremember}[2]{%
    \footnote{#2}
    \newcounter{#1}
    \setcounter{#1}{\value{footnote}}%
}
\newcommand{\footrecall}[1]{%
    \footnotemark[\value{#1}]%
} 

\newcommand{\af}[1]{{#1}}
\newcommand{\cb}[1]{{#1}}
\newcommand{\hh}[1]{{#1}}

\begin{document}

\title{\Large CoronaSurveys: Using Surveys with Indirect Reporting to Estimate the Incidence and Evolution of Epidemics}

\author{
\normalsize Oluwasegun Ojo
\footremember{imdea}{IMDEA Networks Institute, Spain} \footremember{uc3m}{U. Carlos III de Madrid, Spain}
\and \normalsize Augusto García-Agundez\footremember{tud}{TU Darmstadt, Germany}
\and \normalsize Benjamin Girault\footremember{usc}{U. Southern California, USA}
\and \normalsize Harold Hernández\footrecall{uc3m}
\and \normalsize Elisa Cabana\footrecall{imdea}
\and \normalsize Amanda García-García\footrecall{imdea}
\and \normalsize Payman Arabshahi\footremember{uw}{U. of Washington, USA}
\and \normalsize Carlos Baquero\footremember{um}{U. Minho, Portugal} \footremember{inesc}{INESC TEC, Portugal}
\and \normalsize Paolo Casari\footremember{utr}{U. of Trento, Italy}
\and \normalsize Ednaldo José Ferreira\footremember{embrapa}{Embrapa Instrumentation, Brazil}
\and \normalsize Davide Frey\footremember{inria}{Inria Rennes, France}
\and \normalsize Chryssis Georgiou\footremember{ucy}{U. Cyprus, Cyprus}
\and \normalsize Mathieu Goessens\footremember{cf}{Consulting, France}
\and \normalsize Anna Ishchenko\footremember{ntuu}{National Technical U.~of Ukraine ``Igor Sikorsky Kyiv Polytechnic Institute'', Ukraine}
\and \normalsize Ernesto Jiménez\footremember{upm}{U. Politécnica de Madrid, Spain}
\and \normalsize Oleksiy Kebkal\footremember{evo}{EvoLogics, Germany}
\and \normalsize Rosa Lillo\footrecall{uc3m}
\and \normalsize Raquel Menezes\footrecall{um}
\and \normalsize Nicolas Nicolaou\footremember{algo}{Algolysis Ltd, Cyprus}
\and \normalsize Antonio Ortega\footrecall{usc}
\and \normalsize Paul Patras\footremember{edi}{U. Edinburgh, UK}
\and \normalsize Julian C Roberts\footremember{sky}{Skyhaven Media, UK}
\and \normalsize Efstathios Stavrakis\footrecall{algo}
\and \normalsize Yuichi Tanaka\footremember{tuat}{Tokyo University of Agriculture and Technology, Japan}
\and \normalsize Antonio Fernández Anta\footrecall{imdea}
}

\date{}

\maketitle

\begin{abstract}
The world is suffering from a pandemic called COVID-19, caused by the SARS-CoV-2 virus. National governments have problems evaluating the reach of the epidemic, due to having limited resources and tests at their disposal. This problem is especially acute in low and middle-income countries (LMICs). Hence, any simple, cheap and flexible means of evaluating the incidence and evolution of the epidemic in a given country with a reasonable level of accuracy is useful. In this paper, we propose a technique based on (anonymous) surveys in which participants report on the health status of their contacts. This indirect reporting technique, known in the literature as \emph{network scale-up method}, preserves the privacy of the participants and their contacts, and collects information from a larger fraction of the population (as compared to individual surveys). This technique has been deployed in the CoronaSurveys project, which has been collecting reports for the COVID-19 pandemic for more than two months. Results obtained by CoronaSurveys show the power and flexibility of the approach, suggesting that it could be an inexpensive and powerful tool for LMICs.
\end{abstract}


\section{Introduction}

During the current SARS-CoV-2 virus pandemic, monitoring the evolution of COVID-19 cases is very important for authorities to make
informed policy decisions, and for the general public to be informed of the reach of the problem. 
Official numbers of confirmed cases are periodically issued by health authorities \cite{ECDC}. 
Unfortunately, at the early stages of a pandemic outbreak  there is usually only limited ability to test, as well as a lack of other resources. Hence, it is not possible to test all potential cases, and some eligibility criteria is applied to decide who is tested.
Under these circumstances, the official confirmed cases are unlikely to represent the total number of cases (see \cite{maxmen2020much}).
This problem is more pressing in low and middle-income countries (LMICs), which may be limited in their ability to deploy massive testing.
Not having access to reliable data clearly prevents authorities from making informed decisions, putting the population at higher risk.

This motivates the need for other probing techniques, beyond laboratory testing, that can estimate  the number of cases and their evolution. Information obtained using these alternative methods can be more timely, leaving more accurate estimates of the number of cases for later studies  (e.g.,  massive serological studies \cite{yang2012serological}).  Techniques that allow cheap and massive data collection, and lead to reasonably accurate estimates, 
are useful when testing is limited and can lead to improved data-driven decision making. 

Direct surveys are an obvious approach to estimate incidence. A number of these direct surveys to obtain health data have been deployed in various countries in recent months \cite{linares1920estimando,DBLP:journals/corr/abs-2004-01014,FB-survey}. While these surveys can gather  useful data, a large number of participants is needed to achieve reliable estimates. Additionally, these surveys collect sensitive personal health information, which prevents the distribution of the raw data collected, and may also discourage privacy-concerned people from responding.

In this paper, as an alternative to direct surveys, we propose \af{online} surveys with \textit{indirect reporting}, where the questions a participant answers are not about herself, but about her contacts. 
This technique is known in the literature as the \emph{network scale-up method} \cite{bernard1991estimating,bernard2010counting}, and has been successfully used for public health.
This approach has at least two major advantages with respect to direct surveys. First, the survey can be designed so that no personal information is collected from the participant (i.e., it is completely anonymous). Second, indirect reporting has a multiplicative effect, since it reduces the number of responses required to achieve a specific population coverage. The loss in accuracy, due to respondents not always having exact information about the health status of others, may be compensated by the significant increase in coverage (i.e., on average each respondent informs about the likely status of around 100 other \af{people, and this number is rather stable across countries}).

We have designed and deployed a system that implements the network scale-up method via \af{online} anonymous surveys with indirect reporting, as part of the CoronaSurveys project \cite{CoronaSurveys}. This system has surveys in multiple languages and allows reporting data on the incidence of COVID-19 in all countries. \af{The project team has promoted the survey via online social networks and personal contacts. The system} has been collecting data for more than 2 months now, and has collected more than $13,000$ responses, reporting cases in 70 countries. All the collected data is available to be openly used\footnote{\url{https://github.com/GCGImdea/coronasurveys/tree/master/data}}. In parallel with the data collection, the CoronaSurveys project has been developing statistical techniques to estimate the incidence of COVID-19 in different countries and geographical areas.

In the rest of this paper, we present the different elements of the CoronaSurvey project, and compare our resulting estimates with those obtained via other indirect methods and a wide systematic serology study conducted in Spain \cite{ENEcovid19}. Our estimates are surprisingly close to the values obtained in the serological study\footnote{\af{Taking into account the sensitivity of the tests used in the serology study and that there are roughly 34\% asymptomatic cases.}}. From this, we conclude that  
anonymous open indirect surveys, in combination with our proposed statistical techniques, provide a cheap and flexible option for monitoring
epidemics, especially in countries with limited infrastructure.

\section{Estimates via Anonymous Surveys with Indirect Reporting}

This section describes the two main components of the CoronaSurvey project: collection of survey responses and estimation of COVID-19 incidence from the collected responses.

\subsection{Data Collection: The Survey}

Our proposed surveys are carefully designed to avoid querying participants about their own health status\footnote{While in general reporting on one's own health status without any identifying information would not have privacy implications, it could increase the risk for de-anonimization attacks if the raw data is publicly shared, or the survey system is compromised.}, identity, or any personal data. In order to provide data for estimating incidence, participants answer three questions. First, they select a geographical area, which can be a whole country or a region within a country. (The participant does not need to be a resident of that area.)  Then, they answer two simple questions about that area: 
\begin{itemize}
\item How many people do you know in this geographical area? (Please, consider only people whose current health status you likely know.)
\item As far as you know, how many of the above have had symptoms compatible with COVID-19 (or were diagnosed with the disease)?
\end{itemize}

We aim at increasing participation by not asking for any personal information (protecting the participant's privacy), and by having just two questions\footnote{In the new version of the survey we have additional questions to estimate other aspects of the epidemic, but the total number of questions we include remains very small (10 or less), compared with most surveys (e.g., \cite{FB-survey} has more than 30 questions).}. However, the lack of detailed information about the participants makes the estimation process more challenging. In particular, we do not control the spread of the survey and do not have means to ensure that there is an adequate coverage, in terms of regions, age groups and other demographic factors. 

The main novelty of our proposed survey is that a participant does not report on her own health status but on those of others. This typically leads each participant to report on the health status of a large sample (around one hundred on average), which significantly increases our coverage of the population. We believe that this is the reason why, 
somewhat surprisingly, even with the limitations in the available data and few responses, we can still obtain estimates that are less than 4\% away from the real value (see Section~\ref{s-comparison}). 
\af{We believe that indirect reporting compensates significantly the biases in the set of participants.}
Obvious advantages of this approach are that it is very simple to deploy and can give very timely results.

We started the project by running discrete surveys in Spain and evolved the system so that now it collects data continuously.
\af{The survey was initially done via Twitter, but was quickly moved to Google Forms. For privacy reasons, in May 2020 it has been moved to a
dedicated server running the web-based Limesurvey system.}
We have been running the survey (starting with a simpler version) in some countries for more than 3 months (since March 13). The survey is now available in 57 languages. Participants can report at the regional level in 149 countries and at the country level in all countries of the world. We have already collected more than $15,000$ responses for almost 90 countries. \af{This participation has been obtained by advertising and promoting
the survey in online social networks, and via personal contacts. No incentive (economic or otherwise) has been used to promote participation. In some countries (e.g., Spain, Portugal, Ukraine, Brazil) the project received media attention, which led to bursts in participation at specific times. These burst do not seem to influence the results.}

\subsection{Computing Incidence Estimates}

\paragraph{Country-wide and regional estimates} 
We use responses to our survey to estimate the incidence of COVID-19 in different countries. Respondents can report on the number of people they know for a specific region or for the whole country. 
Currently, when the data availability allows it,  estimates for a given country are computed based only on the regional responses, because \af{we observed that} country-wide responses tend to introduce a geographical bias. 

For instance, our initial estimates for Spain were ignoring regional information, and considering all responses as referring to the whole country. However, we observed that most of our responses were actually from the region of Madrid \af{(very likely because the Spanish team members are based in Madrid)}, which is the most affected region in Spain, and thus tended to report high incidence ratios. This was in line with the prevalence in Madrid, but not necessarily a good reflection of country-wide conditions. We then computed new estimates by taking regional information into account and confirmed that we were initially over estimating the number of cases in Spain. 

We also decided to use only the responses that specify a region as it is reasonable to assume that country-wide responses will share a similar bias (coming from the most affected region)\footnote{This probably also results from the fact that we spread our survey through social connections starting from our researchers based in Madrid.}. Note that while we observed this bias in one specific dataset (Spain), similar problems could arise in other countries. COVID-19 outbreaks have been fairly localized in most countries, with different prevalence observed at the region or state level. Thus, answers that provide information within a region tend to lead to more reliable estimates in general. We therefore plan to remove the country-wide option in future survey updates.

\paragraph{COVID-19 incidence estimation -- Region based approach}
Assume the country of interest is divided into $k$ regions. In each region $i$, we get $n_i$ responses, where each response $j$ contains:
\begin{itemize}
    \item a \textbf{reach} variable, $r_i^j$, which is the network size of participant $j$ (i.e., the number of people whose health she knows; answer to the first survey question) and
    \item a \textbf{count} variable, $c_i^j$, which is the number of people (out of the $r_i^j$ reached) that are known to the respondent to be showing symptoms compatible with COVID-19 (answer to the second survey question),
\end{itemize}
for $j \in \{1,\ldots,n_i\}$. Then, we obtain an estimate of the ratio of people infected with symptoms in region $i$, $\hat{p}_i$, using \cite{bernard2010counting}
$$
\hat{p}_i = \frac{\sum_{j = 1}^{n_i} c_i^j}{\sum_{j = 1}^{n_i} r_i^j}.
$$ 

From the ratios $\hat{p}_i$ of the different regions, we compute an estimate of the proportion of those infected with symptoms in the country, $\hat{p}$, using a weighted sum of these ratios as follows.
$$
\hat{p} = \sum_{i=1}^k \omega_i \hat{p}_i,
$$
where $\omega_i$ is a proportional weight defined as 
$$
\omega_i = \frac{N_i}{N},
$$ 
where $N_i$ and $N$ are the populations of region $i$ and the whole country, respectively.  To build a confidence interval for $\hat{p}$, we need to estimate its variance  $V\left( \hat{p} \right)$. To do so, we consider each region as a stratum in a post-stratified random-sampling setting, and estimate the variance of proportions for post-stratified sampling \cite{Holt1979} as 
$$
V\left( \hat{p} \right) = \frac{1-f}{n} \sum_{i=1}^k \omega_iS_i^2 + \frac{1-f}{n^2} \sum_{i=1}^k (1-\omega_i)S_i^2,$$ 
where $n = \sum_i^k n_i$ and $f = n/N$. The value $S_i^2$ for each region can be estimated using
$$S_i^2= \sum_{j=1}^{n_i} \frac{(p_{ij}-\hat{p_i})^2 }{n_i-1},$$
where $p_{ij} =  \frac{c_{i}^j}{r_i^j}$. We can then construct a confidence interval for $\hat{p}$ as
$$\hat{p} \pm 1.96 \sqrt{V(\hat{p})}.$$

Before generating an estimate for a given country at a given date, we first clean the survey responses by identifying and removing outliers. We declare a response to be an outlier if  $r_i^j$, the number of persons that the participant claims to know, is unusually large (specifically, we remove entries where  $r_i^j$ is beyond 1.5 times the interquartile range above the upper quartile). We also consider to be outliers responses leading to a large ratio $p_{ij}$ of symptomatic people reported (specifically, we remove entries in which the ratio $c_{i}^j/r_i^j$ is above $0.3$). We remove responses with large ratios because we aim at surveying the general population, and not individuals (such as doctors or nurses) who may be in contact with a large number of symptomatic cases. 

After outlier removal, for any given day we aggregate data from that day and from previous days until we have at least $A_{\min}$ responses. For the experiments we report here we set $A_{\min} = 300$, as we observed  empirically that this provides enough data to make our estimate reliable. Since we usually do not get $A_{\min}$ responses on a given day for most countries, using data from previous days also provides a rolling estimate, which has an implicit smoothing effect on the estimate. From these responses, we exclude country-wide responses (as discussed earlier) and then compute the value of $\hat{p}_i$ for each region $i$ represented in the responses. Then, the estimate of the proportion of people infected with symptoms in the country, $\hat{p}$, for the day is computed as shown above.

The described procedure produces reasonable estimates 
and works well as long as we have a sufficiently large number of responses (per day).  
Hence, we have only applied this procedure to generate estimates from data gathered in Spain, Portugal, and Ukraine, countries from which we got the highest number of responses. 

\paragraph{COVID-19 incidence estimation -- Country based approach}
For countries where the number of survey responses is smaller (after removing outliers), we have used a simpler estimation procedure. Denote $n_d$ the number of responses collected on day $d$ for the given country (counting both regional and country-wide responses after removing outliers). If $n_d \geq a_{\min}$, we estimate the incidence of COVID-19 in the country of interest using
$$
\hat{p}_d = \frac{\sum_{l = 1}^{n_d} c_l}{\sum_{l = 1}^{n_d} r_l},
$$
where $r_l$ is the number of people a participant $l$ declares to know in the first question of the survey, and $c_l$ is the number of people (out of $r_l$) showing symptoms compatible with COVID-19, for $l \in \{1,\ldots, n_d\}$. On the other hand, if $n_d < a_{\min}$, we do not compute an estimate for day $d$. We instead aggregate the responses for day $d$ to the responses for the subsequent days $d+1, d+2, \ldots$ until we have a day $d'=d+m$ such that $n_m = n_d + n_{d+1} + \ldots + n_{d+m} \geq a_{\min}$ responses. In the experiments reported here we empirically chose $a_{\min}=30$.  We then compute the estimate $\hat{p}_{d'}$ for day $d'$ as\footnote{We expect that better results may be achievable if $A_{\min}$ and $a_{\min}$ are selected as a function of country population. We plan to investigate this in future work.}

$$\hat{p}_{d'} = \frac{\sum_{l = 1}^{n_m} c_l}{\sum_{l = 1}^{n_m} r_l}.$$

The estimates $\hat{p}_d$ and $\hat{p}_{d'}$ are simple proportions and we construct a 95\% confidence interval for $\hat{p} \in \{\hat{p}_d, \hat{p}_{d_m}\}$ using confidence intervals for binomial proportions as follows
$$\hat{p} \pm 1.96\sqrt{\frac{\hat{p}(1-\hat{p})}{r}},$$
where $r = \sum_{l} r_l$. A major limitation of this method is that we do not obtain estimates for every day. However, we are constantly refining these techniques, and trying new ways to obtain better estimates. 

\section{Validation}

In the previous sections we introduced our strategy for indirect reporting of the number of cases. Given the nature of an open and anonymous survey, where data quality is much harder to enforce, we soon identified the need for an independent estimator to which we could compare the survey results. In most countries we had access to time series with the number of official, RT-PCR confirmed, cases and COVID-19 mortality data. Together these series allow the derivation of a naive \emph{case fatality ratio} (CFR). 

Up to early May 2020 our running option for an independent estimator was to use the CFR to estimate the current number of cases in each country, as detailed in the next section. On May 13, 2020, a large-scale serology study was reported for Spain \cite{ENEcovid19} and this provided a more precise and direct data collection for calibration. 

\subsection{Inferring Cases from Reported Mortality}

In an ongoing epidemic, the current CFR should be calculated by taking into account the number of deaths \hh{($d$)} over the number of current cases with known outcomes \hh{($c$)}, since very recent cases can still evolve as fatalities or recover \cite{nishiura2009early}. This correction yields a corrected CFR (cCFR), more accurate than the naive CFR that is often reported. \cb{To perform the correction we follow the same methodology that is described in \cite{russel2020using} as well as their code in \url{https://github.com/thimotei/CFR_calculation}. We keep the same estimates for the delay from case confirmation to death, i.e., a Lognormal distribution with a mean delay of 13 days and a standard deviation of 12.7 days. (Our code is available under \url{https://github.com/GCGImdea/coronasurveys}.)}

Under the assumption that the disease has similar mortality rates for similar populations, it is possible to use stable $\textit{cCFR}$ estimates to obtain a baseline, $\textit{cCFR}_b$, for COVID-19 and check how each country's current $\textit{cCFR}$ compares to that baseline, and hence infer the proportion of cases that are being detected. \cb{In \cite{russel2020using} the authors keep a frequently updated estimate of the level of under-reporting for several countries. As an example if $\textit{cCFR}_b=1\%$ and a country exhibits a $\textit{cCFR}=2\%$, they infer that only $50\%$ of the cases are being detected.}   
\cb{For our purposes, if we multiply the reported number of cases by $\frac{\textit{cCFR}}{\textit{cCFR}_b}$, we obtain an estimate for the likely true number of cases in that given country. In the example above we would have multiplied the reported number of cases by $\frac{\textit{cCFR}}{\textit{cCFR}_b}=2$.}

\cb{The next step is to select a reference baseline}. In \cite{Verity2020} the authors report, from a large sample of cases in China, a baseline $\textit{cCFR}_b$ of $1.38\%$ \hh{($d_b = 1,023; \mbox{ and } c_b = 74,130$)}, \cb{and we have chosen to use this baseline in our estimators.} Still, given that several countries (Korea, New Zealand, and others) have stabilized the growth of COVID-19 cases, it is possible to use the data in those countries to also define a baseline.
\hh{Following the Ln-Method in \cite{10.2307/2531405}, we also construct a confidence interval for the ratio $\frac{\textit{cCFR}}{\textit{cCFR}_b}$, at any given date. We model the number of deaths $d$ using a binomial distribution with parameters $\textit{cCFR}$ and $c$ (analogously, we refer to $d_b, \textit{cCFR}_b$, and $c_b$ for our fixed baseline). Then, it can be shown that $\ln \left( \frac{d/c}{d_b/c_b} \right) $ is approximately normally distributed with mean $\ln \left( \frac{\textit{cCFR}}{\textit{cCFR}_b} \right)$ and estimated variance $\hat{\sigma}^2=1/d-1/c+1/d_b-1/c_b$. Finally, a 95\% CI for our ratio is given by $\left( r \cdot \exp(-1.96 \cdot \hat \sigma ),\; r \cdot \exp(1.96 \cdot \hat \sigma ) \right)$, where $r$ is the observed value of $\frac{d/c}{d_b/c_b}$. }

We observe that one limitation of resorting to reported mortality data is that some countries might not properly report or classify it as COVID-19 mortality. However, in countries with adequate reporting, it is a useful source of calibration and independent estimation.



\subsection{Serology Study in Spain}

On May 13, 2020 the Spanish Government published a first report from a large-scale serology study that looked for COVID-19 antibodies in the population \cite{ENEcovid19}. Samples were collected from April 27 to May 11, and results made available from a group of $60,897$ participants in the study, selected according to demographic criteria, to obtain a representative sample of the population. Although both IgM and IgG antibodies were measured, the report focuses on the prevalence of  SARS-Cov2 IgG antibodies. The overall prevalence was reported as 5.0\% (95\% CI: 4.7\%-5.4\%), and regional variations ranged from 1.1\% in Ceuta to 11.3\% in the \emph{Comunidad de Madrid}.

The IgG test was found to have a sensitivity of 79\% and specificity of 100\%. Given that, we can correct for false negatives and estimate an overall infection rate of 6.33\% (=0.05/0.79), \hh{with approximate 95\% CI: 5.95\%-6.84\%}. Assuming an average time span of two weeks since infection to the development of detectable levels of IgG \cite{long2020antibody}, and with a population in Spain of $46,934,628$ persons, this leads to an estimate of approximately $2,970,546$ \hh{(95\% CI: 2,792,313-3,208,190)} cumulative cases around the weeks from April 13 to 27 (in the middle of this period, on April 20, the number of RT-PCR confirmed cases was $200,210$, almost 15 times lower). Using the cumulative mortality on May 11 (roughly two weeks after the likely infection dates) of $26,744$, this leads to an estimated \emph{infection mortality rate} (IFR) of 0.9\% \hh{(95\% CI: 0.83\%-0.96\%)}\footnote{\hh{The data of confirmed cases and mortality was extracted from \cite{ECDC} on the date of the submission (May 20, 2020).}}. \cb{In Brasil, another large-scale serology survey (with 25,025 participants) \cite{Hallal2020.05.30.20117531} recently provided an estimated IFR of 1\%, a value that is in line with our estimate for Spain.}



The Spanish study includes data on the proportions of IgG positives that had one or more symptoms, or were completely asymptomatic. The proportion of IgG positives with some kind of symptoms (i.e., at least one symptom) \hh{was reported to be 66.27\%. Fixing this percentage, and assuming that most of the (officially reported) RT-PCR confirmed cases were symptomatic}, one can estimate about $1,968,550$ persons with symptoms, and a CFR of around 1.36\% (actually, very close to the value calculated for Wuhan at 1.38\% \cite{Verity2020}).

\begin{figure}[htb]
\begin{center}
\includegraphics[width=0.9\linewidth]{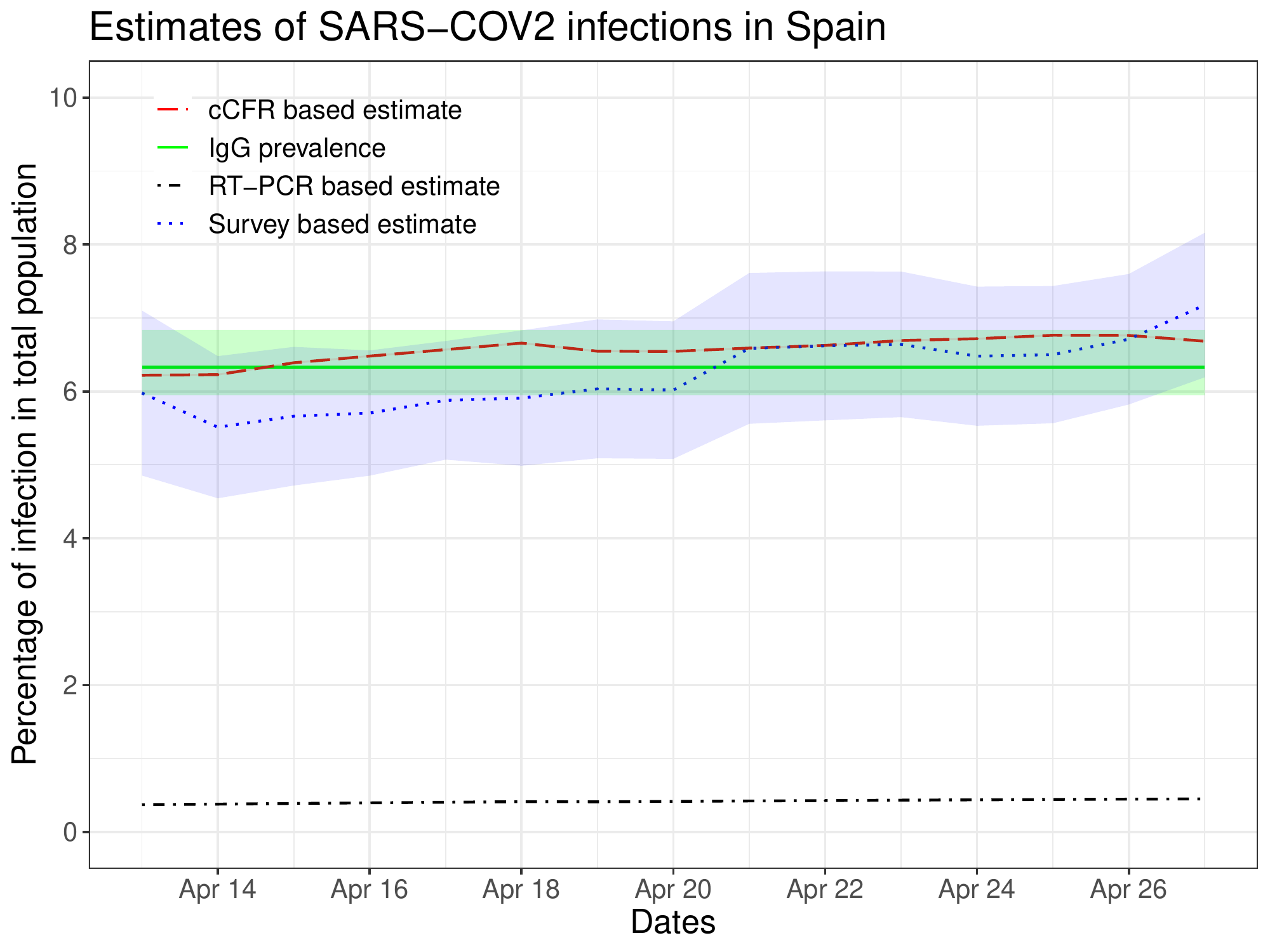}
\end{center}
\caption{\hh{Case estimates with 95\% confidence bands for Spain, April 13 to 27, 2020.}}
\label{cal}
\end{figure}

\begin{figure}[htb]
\begin{center}
\includegraphics[width=1.0\linewidth]{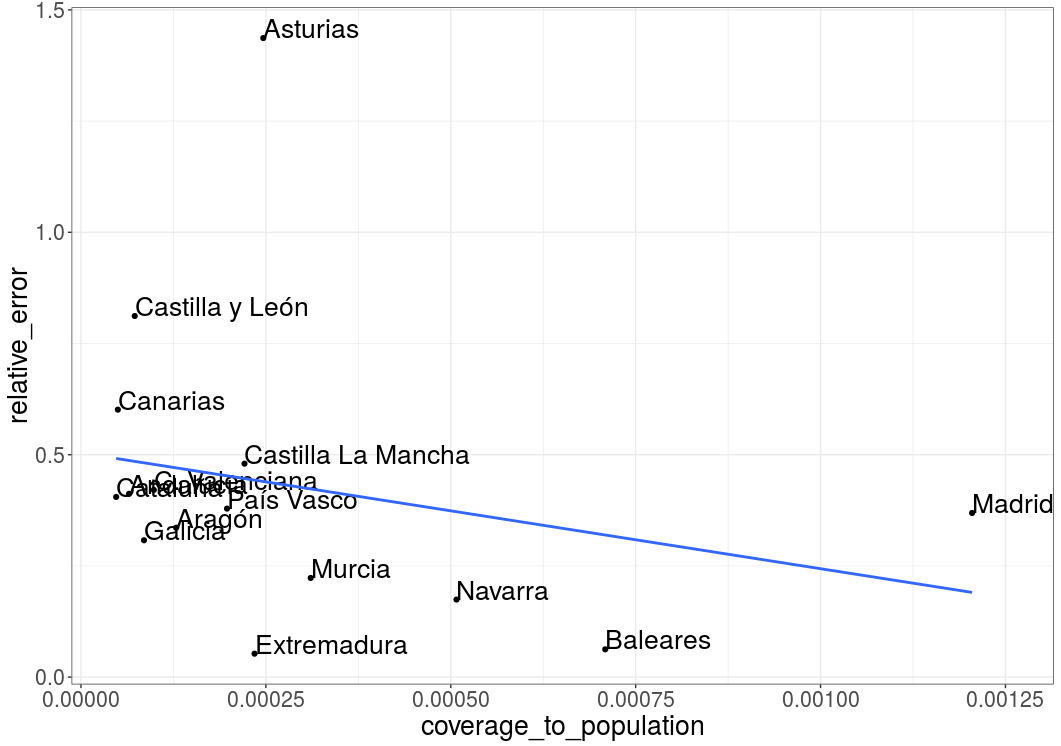} 
\end{center}
\caption{Relative accuracy vs.~local coverage: For each region, we plot the relative error in our estimates as compared to serology data, shown as a function of the relative reach within the region. Higher reach (better coverage within a region) tends to lead to lower error. }
\label{fig:coverage-vs-accuracy}
\end{figure}

\subsection{Comparing Estimates}
\label{s-comparison}

Adopting as ground truth the serology-derived value of 6.33\% cumulative infections in the period from April 13 to 27 (which lead to 79\% IgG positives two weeks later, i.e. 5\%), we compare this ground-truth to our estimates in the same period. Since our estimates, cCFR-based and survey-based, target symptomatic cases, we need to scale them to total infections by dividing the estimates by 0.66 (the ratio of symptomatic over total infections as reported in the study \cite{ENEcovid19}). We also show, for comparison, the number of reported RT-PCR confirmed cases.

As we can see in Figure \ref{cal} the official RT-PCR data is about an order of magnitude less than the likely true rate of infected people that is inferred from IgG prevalence, 6.33\%. Figure \ref{cal} also shows that both the cCFR-based estimate and the estimate derived from the open survey closely track the IgG-prevalence value. \hh{Furthermore, the confidence bands intersect throughout the period, with a clear difference in amplitude due to the sample sizes}.  
The average of the cCFR estimates in the 15-day period was 6.56\%, while that of the survey-based estimates was 6.2\%. This places them, respectively, only 0.24\% and 0.13\% apart from the IgG reference value (average relative difference of 3.72\% and 2.05\%, respectively). These results show that open surveys can bring relevant data on the size of a pandemic, which can be be useful when more reliable metrics are not yet available or better estimates cannot be implemented in some regions.

For further evaluation of our method we compare our estimates to the serology data for each region in Spain as shown in Figure \ref{fig:coverage-vs-accuracy}. We plot the relative error in our estimate for a region as a function of the relative coverage of our surveys for that region (reach divided by population). Although we can see some variability, we note that the trend is towards lower relative error as the reach increases.

\section{Experience Using Open Surveys in LMI Countries}

We have obtained estimates for a number of countries. In particular, we have received enough responses to estimate symptomatic cases in 3 LMICs, namely Brazil, Ecuador, and Ukraine (see Tables \ref{tab:LMICs1} and \ref{tab:LMICs2}). 

\begin{table*}
    \centering
    \footnotesize
    \begin{tabular}{|l|c|c|c|c|c|c|}
    \hline
 Country    &  Date       & Cases & Fatal. & cCFR & CoronaSurveys \\ 
            &              &   &  & (CI 95\%)  & (CI 95\%)  \\
 \hline
  Brazil     & May 17 & $233,142$ & $15,633$ & $2,139,681$ & $2,120,134$ \\
     & & & & ($2,135,408$-$2,143,963$) & ($1,195,676$-$3,044,593$) \\
  Ecuador    &  April 15 & $7,603$ & $355$ & $53,435$ & $274,668$ \\
       & & & & ($53,069$-$53,804$) & ($190,236$-$359,100$)  \\
   Ukraine    & April 26 & $8,617$ & $209$ & $32,078$ & $246,646$  \\
        & & & & ($31,734$-$32,426$) & ($107,482$-$385,811$)  \\ \hline
    \end{tabular}
    \caption{Summary of estimates for Brazil, Ecuador and Ukraine. Cases and Fatalities correspond to the official data on that day, while cCFR and CoronaSurveys are estimates. }
    \label{tab:LMICs1}
\end{table*}

\begin{table}[t!]
    \centering
    \footnotesize
    \begin{tabular}{|l|c|c|c|c|c|c|}
    \hline
 Country    &  Date       & Cases & cCFR & Resp. & CoronaSurveys \\  
            &         &   \% pop.    & \% pop. (CI 95\%) &            & \% pop. (CI 95\%)  \\
 \hline
  Brazil     & May 17 & $0.11\%$ & $1.01\%$  ($1.00$ - $1.01$) & 41  & $1.00\%$ ($0.56$ - $1.43$)  \\
  Ecuador    & April 15 & $0.04\%$ & $0.31\%$  ($0.31$ - $0.31$) & 30  & $1.61\%$ ($1.11$ - $2.10$)  \\
   Ukraine    & April 26 & $0.02\%$ & $0.07\%$  ($0.07$ - $0.07$) & 30  & $0.56\%$ ($0.25$ - $0.88$)  \\ \hline
    \end{tabular}
    \caption{Summary of estimates for Brazil, Ecuador and Ukraine in percentage of the country population.} 
    \label{tab:LMICs2}
\end{table}

In Brazil we have recent estimates for May 17, when the official number of cumulative confirmed cases 
was $233,142$, and the official number of cumulative fatalities was $15,633$. For that same date, the estimate based on the cCFR is $2,139,681$ (CI 95\%: $2,135,408$ - $2,143,963$), 1.01\% of the population, and the estimate based on 41 survey responses is $2,120,134$ (CI  95\%: $1,195,676$ - $3,044,593$), 1\% of the population. Not surprisingly, the estimated number of cases is one order of magnitude larger than the number of confirmed cases. However, it is remarkable how close the cCFR and survey-based estimates are to each other, differing in less than $20,000$ cases (0.01\% of the population).

In Ecuador we also observe that the number of estimated cases is at least one order of magnitude larger than the official number of confirmed cases. However, we observe that the estimates from cCFR and from the surveys are also very different. For instance, we have estimates dated April 15, when the official number of confirmed cases was $7,603$ and the official number of fatalities was $355$. Our case estimate for that date from the cCFR is $53,435$ (CI $53069$ - $53804$), 0.31\% of the population, and the one from 30 survey responses is $274,668$ (CI $190,236$ - $359,100$), 1.61\% of the population. We observe a significant difference between the two estimates, the one from the surveys being 5 times larger than the cCFR-based one. This difference does not seem to be the result of geographical bias, since few survey responses came from the provinces with the largest number of cases (e.g., Guayas).

We observed a similar behavior in Ukraine where, again, the estimates are at least one order of magnitude larger than the number of confirmed cases, and the number of estimated cases from the surveys is one order of magnitude larger than the estimates from cCFR. For instance,
the latest direct estimate from (30) survey responses was done on April 26, and has a value of $246,646$ (CI $107,482$ - $385,811$), 0.56\% of the population, while the cCFR estimate is $32,078$ (CI $31,734$ - $32,426$), 0.07\% of the population. The confirmed numbers of cases and fatalities on that date were $8,617$ and $209$, respectively. This result is not due to a geographical bias, since the region-based estimate for that same day from 300 survey responses is $159,529$ (CI $62,361$ - $256,696$), 0.35\% of the population, which is lower, but still five times higher than the cCFR-based estimate.

The results in Ecuador and Ukraine are puzzling, and we are not able to explain them yet. Our current hypothesis is that these countries use different criteria for reporting cases and fatalities than the countries we use as reference. This may cause the cCFR-based estimate to be unreliable. We have deployed additional questions in the survey that we believe can be used to track the difference. The good news is that this can be done with very little effort, and we do not need a lot of responses to have enough information to have solid conjectures.

\section{Discussion}

By now it is clear that relying only on confirmed cases and fatalities to measure the true size of a growing pandemic is not a good idea. It is possible to use this data to derive estimates, like the one we obtain here based on the cCFR, that are
reasonably reliable in countries with a good reporting system. This has been shown in the case of Spain using the ground
truth provided by a serological study. The same study indicates that open anonymous surveys with indirect reporting also provide estimates that are close to the real values. The matching between cCFR-based and survey-based estimates has also been observed in Brazil. However, we have found countries in which there is no ground truth, and the cCFR-based and survey-based estimates differ significantly. We are investigating this further by adding new questions to the surveys in these countries,
which will provide additional information on the causes of these discrepancies.

We are aware that having open anonymous surveys prevents a tight control of the population of participants. Hence, our responses may be suffering from strong biases: for example, they may be close geographically and socially to the team members, which are the ones promoting the survey. We are also aware that the set of people and cases participants report are possibly not disjoint. Interestingly, until now the only bias that we have observed to be relevant is the geographical bias. \af{Initial studies done via simulation hint that the} intersection of contacts among participants does not seem to have much influence on the estimates. \af{However, more experiments are required. In fact, new questions have been added to the survey to evaluate the level of overlapping, and estimate the propagation graph of participation.}

However, we plan to explore ways to influence \af{and select} the population of participants. One line to follow is the use of targeted campaigns in social networks and web ads (using Facebook Ads or Google Ads), in which we have control over who sees a given ad promoting the survey. This will allow targeting participants from certain geographical areas and certain demographic profiles. \af{In fact, we have started a exploratory campaign using Facebook Ads in Brazil.}

While we only present here techniques for obtaining estimates on the number of people infected with COVID-19, we are designing new surveys that will hopefully allow us to estimate other important parameters of the pandemic, like the number of newly infected cases, the reproduction number, or the forecast needs of health equipment and infrastructures.

We believe that a survey system like the CoronaSurveys project is especially suited for LMICs, since the cost of preparing and deploying a survey is extremely small, participants can use very simple devices to fill the survey (since it is web based), and the number of participants required to have information on the pandemic is rather low. Having a uniform approach to obtain this information in many countries also makes the process especially interesting, since it avoids the current problems with different ways of counting and measuring.

\section{Acknowledgements}

We would like to thank the large group of researchers and collaborators that is currently involved in the CoronaSurveys project in any form: 
Annette Bieniusa,
Ignacio Castro, 
Chus Fernández,
Angeliki Gazi, 
Rodrigo Irarrazaval,
Alvaro Méndez,
Esteban Moro,
Paul Rimba,
Erol Sahin,
Andres Schafer,
Ghadi Sebaali,
Natalie Soto,
Matias Spatz Fernández,
Christopher Thraves, 
Pelayo Vallina, 
Lin Wang. Especial thanks to the Crowdfight COVID-19 project \cite{crowdfight} and its volunteers.

\bibliographystyle{plain}
\bibliography{refs}

\end{document}